\documentclass[aps,prl,nofootinbib,twocolumn,superscriptaddress]{revtex4}

\usepackage{graphicx}

\newcommand{\beq}{\begin{equation}}
\newcommand{\eeq}{\end{equation}}

\newcommand{\bea}{\begin{eqnarray}}
\newcommand{\eea}{\end{eqnarray}}

\newcommand{\gsim}{\lower.7ex\hbox{$\;\stackrel{\textstyle>}{\sim}\;$}}
\newcommand{\lsim}{\lower.7ex\hbox{$\;\stackrel{\textstyle<}{\sim}\;$}}

\def\mysection#1{{\bf #1.} }

\begin{document}

\title{Double gamma-ray lines from unassociated Fermi-LAT  sources revisited
 }

\author{Andi Hektori}
\email[]{andi.hektor@cern.ch}
\affiliation{NICPB, Ravala 10, 10143 Tallinn, Estonia}
\author{Martti Raidal}
\email[]{martti.raidal@cern.ch}
\affiliation{NICPB, Ravala 10, 10143 Tallinn, Estonia}
\affiliation{CERN, Theory Division, CH-1211 Geneva 23, Switzerland}
\affiliation{Institute of Physics, University of Tartu, Estonia}
\author{Elmo Tempel}
\email[]{elmo@aai.ee}
\affiliation{NICPB, Ravala 10, 10143 Tallinn, Estonia}
\affiliation{Tartu Observatory, Observatooriumi 1, T\~oravere 61602, Estonia}

\date{\today}

\begin{abstract}
We search for the presence of double gamma-ray line from unassociated Fermi-LAT sources including detailed Monte Carlo simulations to
study its global statistical significance. Applying the Su \& Finkbeiner  selection criteria for high-energy photons we obtain a similar excess over 
the power-law background from 12 unassociated sources. However, the Fermi-LAT energy resolution and the present low statistics does not allow to 
distinguish a double peak from a single one with any meaningful statistical significance. We study the statistical significance of the fit to data with Monte Carlo 
simulations and show that the fit agrees almost perfectly with the expectations from random scan over the sky.   
We conclude that the claimed high-energy gamma-ray excess over the power-law background from unassociated sources is nothing but 
an artifact of the applied selection criteria and no preference to any excess can be claimed with the present statistics.

\end{abstract}


\maketitle

\mysection{Introduction}
The evidence for 130~GeV gamma-ray excess from Galactic centre consistent with dark matter (DM) annihilations in the central cusp
is by now well established fact~\cite{Weniger:2012tx,Tempel:2012ey,Su:2012ft}. Therefore searches for  the signal of DM annihilations from
other DM dominated objects is of utmost importance. An evidence for the same 130~GeV peak has been found from nearby galaxy
clusters supporting the DM annihilations as an origin of the peak~\cite{Hektor:2012kc}. The two signals come from known DM dominated
sources. A search for possible unidentified sources of the 130~GeV gamma-ray peak was performed in~\cite{Tempel:2012ey} where
several of them have been identified. It is attempting to associate this signal with DM subhaloes of our  Galaxy. 
However, due to large ``look elsewhere effect" related to unidentified sources those can just be statistical upward fluctuations of the 
background~\cite{Boyarsky:2012ca}. More data is needed to resolve this issue.

Recently a similar search for possible gamma-ray excess from unassociated Fermi-LAT point-like sources was performed in~\cite{Su:2012zg}.
The motivation for that search was a claim that possibly those sources can be identified with DM subhaloes, exactly the same motivation 
as in~\cite{Tempel:2012ey}. The smart trick the authors use is to search for a double peak~\cite{Cline:2012nw} motivated by most of particle physics models of
DM annihilations to photons. If such a double peak is observed, this would favour particle physics origin of the excess over astrophysics.
Although the present low Fermi-LAT statistics together with its limited energy resolution does not allow to distinguish
a double peak structure from a single broad peak with a meaningful statistical significance~\cite{Rajaraman:2012db}, 
the double peak gives marginally  better fit to Galactic centre data~\cite{Su:2012ft}. 
The authors of \cite{Su:2012zg} claim to have observed a double peak with $3.3\sigma$ local statistical significance over the power-law background.

The analyses and results of  \cite{Su:2012zg} can be criticized based on two arguments. First, the Fermi unassociated sources are identified
by gamma-rays with much lower energy than the 130~GeV peak. Consequently the first question to ask is  why those sources 
play any role at high-energies and what is the possible connection between the low and high-energy gamma-rays. Those issues have been addressed 
in~\cite{Hooper:2012,mirabal}. Second, the authors of  \cite{Su:2012zg}  make no attempt to estimate the global significance of their claim. 
While stacking together data from several regions, they should have estimated the expected effect of their selection procedure by
choosing arbitrarily the same number of regions in the sky, and scan over the full area as is done in~\cite{Hektor:2012kc}.
This is the issue addressed in this note.

 The aim of this work is first to repeat the analyses of  \cite{Su:2012zg} using their criteria for selecting high-energy photons, 
 and then to apply the  Monte Carlo simulations similar to the one performed in \cite{Hektor:2012kc} to study  their claim. 
We do find a similar excess of high-energy gamma-rays over the power-law background 
as in \cite{Su:2012zg} with a similar local significance confirming that the 
idea of looking for the double peak over the power-law background is indeed very useful one.  
However, we show that with Fermi-LAT energy resolution and with present 
{\it very} limited statistics one cannot distinguish between the double and single peaks with any meaningful significance. 
We do not observe any systematic effects in the signal that could be coursed by the detector since those must have 
occurred everywhere in the data, including background. Thus the claims of detector effects in ~\cite{Hooper:2012,mirabal} 
have no justification at present. Performing the Monte Carlo study to determine the
global significance of the excess we find that the selection criterion of \cite{Su:2012zg} itself implies a peak, and the peak from the
unassociated Fermi sources agrees almost perfectly with the Monte Carlo expectation.
 We conclude that the claimed excess is nothing but an artifact due to the applied event selection and is not associated 
 with the Fermi-LAT point-like sources. Thus the global
 statistical significance of the excess is completely negligible and the results of  \cite{Su:2012zg} are systematically biased.

\begin{figure}[htbp]
\begin{center}
\includegraphics[width=0.47\textwidth]{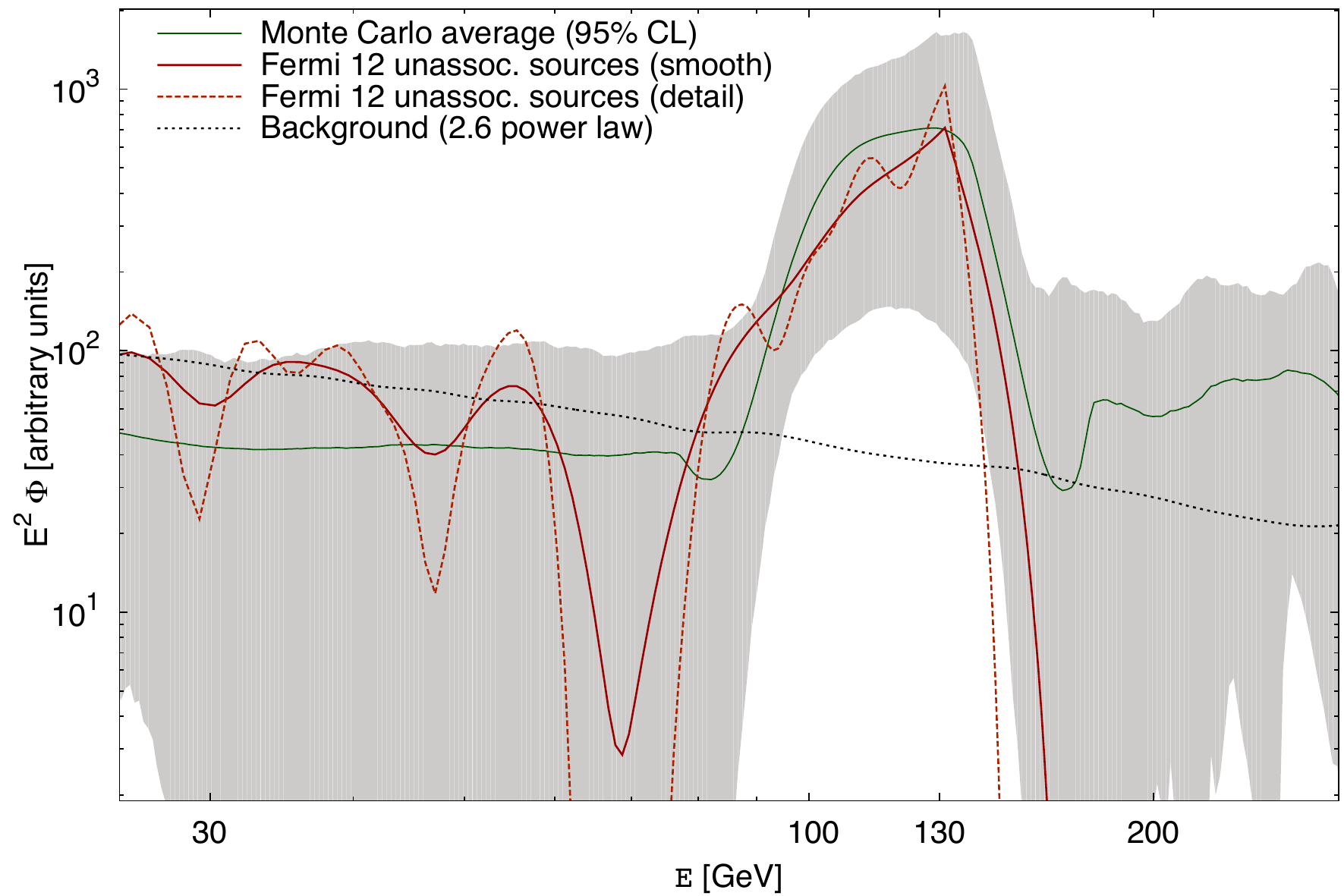}
\caption{Fit to Fermi-LAT gamma ray spectrum from 12 unassociated sources selected as described in the text (solid red curve) 
together with the average Monte Carlo expectation from 12 arbitrary regions in the sky selected by the same criteria (solid green curve).  
The $2\sigma$ error-band obtained with Monte Carlo technique is presented by the grey band. 
Fit to data with unphysical energy resolution (narrow kernel) is also presented with dashed red line.
 The excess claimed in Ref.~\cite{Su:2012zg} is clearly an artifact of the selection criteria of photons and can be obtained for arbitrary selection in the sky. } 
\label{fig1}
\end{center}
\end{figure}

\mysection{Data analyses and Monte Carlo results}
The authors of \cite{Su:2012zg} work with 319 unassociated Fermi-LAT sources and select a subset of them (16 sources) which have at least one photon 
in the energy range 100-140~GeV. We find 12 objects like that using the data selection constraints as described in \cite{Tempel:2012ey} and suggested by Fermi-LAT team, which is 4 less than found in \cite{Su:2012zg} (but in agreement with~\cite{Hooper:2012}). 
We note that the total number of photons in the energy range 20-300~GeV
from those 12 sources is  37 that is really small number for any meaningful statistics. 
For comparison, in our work  \cite{Tempel:2012ey} we did not consider any individual sources with less than 80 high-energy photons at all. 
Nevertheless, stacking together photons from the 12 sources  we observe an excess over power-law background depicted with continuos 
red line in Fig.~\ref{fig1}. This is very similar to the result obtained in  \cite{Su:2012zg}. 
Closer look at these 12 sources reveal that there are only 12 photons in the energy range 100-140~GeV.
 Interestingly, 9 of them are either close to the 111~GeV (108.9--116.6~GeV) or  129~GeV (124.7--133.4~GeV) energy
 intervals.  We estimate that the probability of this happening is rather low (1.5\%).  
 This result is the reason behind the claims of \cite{Su:2012zg} observing the double line with high local statistical significance.

However, notice that the fit of data in Fig.~\ref{fig1} (the red continuous line) does not posses a double peak. The reason is that 
physically meaningful kernel size for this analyses is determined by the Fermi-LAT energy resolution. We remind for the reader that
our analyses is based on kernel smoothing method that is independent of binning and refer the reader to Ref.~\cite{Tempel:2012ey}
for technical details. If we choose narrower kernel we can obtain the double peak structure as depicted in  Fig.~\ref{fig1} with red dashed line.
However, such a choice is not physically well motivated at the moment because of limited  Fermi-LAT energy resolution. 
Playing with the existing data (choosing  small kernel sizes in our case or choosing binning in the case of  \cite{Su:2012zg})
one can, indeed, see some evidence for the double peak. However,
its statistical significance over a singe peak is just marginally better and the present Fermi-LAT energy resolution together with low statistics does not allow to 
distinguish the double peak from a single one.
This criticism is completely general and applicable to all searches for the double peak in Fermi-LAT data. 
Only the future experiments with high energy resolution and high statistics can really tell the two peaks apart.

Coming to the main motivation of this work,  a selection of 12 unassociated sources (that are identified by low-energy gamma rays)
with one energetic photon  cannot be physically justified. To the contrary,  we expect that such a procedure is just choosing the expected signal out of 319 samples by hand.  Repeating the same with any 12 energetic photons in the sky should give the similar peak. 
Thus we expect such a ``signal" to be an artifact of the selection of photons and not to be a signal associated with the particular objects.
The authors of \cite{Su:2012zg} do not make an attempt to estimate statistical significance of the excess using Monte Carlo methods 
by scanning over the sky as we did in \cite{Hektor:2012kc}.
Instead they limit themselves to the 16 point-like sources. This is not a valid statistical sample and  cannot be used for estimating 
the statistical significance of the excess.

To show that our criticism is well justified we use the selection criteria of \cite{Su:2012zg} and repeat their analyses for 12 regions in the sky using Monte Carlo method described in   \cite{Hektor:2012kc}.
The results are presented in Fig.~\ref{fig1}. The green line is an average of Monte Carlo result, this is what one typically expects
to obtain if one repeats the procedure of \cite{Su:2012zg}  for any 12 arbitrary regions in the sky. This agrees surprisingly well with the
fit to data of 12 unassociated sources. In fact, by comparison of the red and green lines one can conclude that the data from 12 unassociated sources
follows the Monte Carlo expectations almost perfectly. The $2\sigma$ error band of the Monte Carlo result is presented in Fig.~\ref{fig1}
with the grey band. The data is well within the band.  We conclude that the observed excess has nothing to do with the 
point like sources, this is just an artifact of the procedure of choosing the photons.
Therefore the statistical significance of the excess claimed in \cite{Su:2012zg}  is clearly and systematically overestimated. 
In fact, our results show that no excess can be claimed at all.

For completeness we also studied low-energy gamma-ray spectra of the 12 unassociated sources and find their spectral shapes 
to be all very different from each other.
If the low energy photon spectrum in those regions is dominated by the continuum spectrum of the $Z\gamma,$ $h\gamma$ final states of DM
annihilations, the low-energy spectra must also be identical. We conclude, in agreement with \cite{Hooper:2012,mirabal}, that 
this is not the case.

We finally comment that the 16 candidate sources found in \cite{Su:2012zg} are mostly 
located close to the galactic disc, exactly as the sources in \cite{Tempel:2012ey}. 
Whether those are DM subhaloes or not requires independent
checks. The present poor  statistics does not to allow for definite conclusion.

\mysection{Discussion and conclusions}
The claim that high-energy photons from Fermi unassociated sources originate from DM annihilations in DM subhaloes of our Galaxy
has been criticized in \cite{Hooper:2012,mirabal} based on the shape and magnitude of the low energy photon signal from the selected 12
unidentified sources. Both those papers show that at most 2 sources out of 12 can be potentially associated with DM subhaloes, 
and most probably the low-energy signal comes from AGNs~\cite{mirabal}.
While this criticism might be relevant, it is based on the model dependent assumption that the high-energy and low-energy photons
are physically associated. For example, it is logically possible that the Fermi unassociated sources are, indeed, misidentified astronomical sources
which are within DM subhaloes.  

To explain the findings of \cite{Su:2012zg} the authors of  \cite{Hooper:2012,mirabal} have also argued that the double peak in 
unassociated sources is potentially due to some  kind of unknown artifact of Fermi-LAT detector without showing what this 
artifact could be or how it would give the double peak. Moreover, from Monte Carlo simulations (from arbitrary regions in the sky) we do not observe this double peak, showing that the double peak is not visible everywhere as would be expected if this is unknown artifact of Fermi-LAT detector.

Since the low energy data cannot rule out the possibility that high-energy signal comes from DM subhaloes, we have studied the 
claims related to the high-energy photon signal in detail. We find that choosing the 12 Fermi unassociated sources using the selection criteria for high-energy
gamma-rays as in \cite{Su:2012zg}, stacked data, indeed, shows a peak over the power-law background. 
For the 12 sources there are only 12 photons in the energy range 100-140~GeV, and 9 of them are close to either the 111~GeV  
or the 129~GeV  energies. Probability for this to happen is less than 2\% (according to Monte Carlo simulations considering only the high-energy range 100-140~GeV),  and this is the reason why  \cite{Su:2012zg} 
claim the double peak in data.  However, we show that the energy resolution of Fermi LAT, together with poor statistics, does not allow to 
to distinguish the double peak from one peak with any meaningful significance, as  demonstrated in Fig.~\ref{fig1}.
At present this result applies to any search for double peaks in Fermi LAT data.

The most important result of this work is that we have demonstrated with Monte Carlo simulations that the excess from the 12 unassociated 
Fermi-LAT sources is in a perfect agreement with expectations if one scans randomly over sky and chooses 12 energetic photons according to the 
criteria of \cite{Su:2012zg}. Thus the excess over power-law background is an artifact of selecting high-energy photons out of the signal sample
and has nothing to do with the unassociated sources.  Therefore the choice of 16 point-like sources as an statistical ensemble,
as done in \cite{Su:2012zg}, greatly overestimates
the statistical significance of the excess. Since the fit to data agrees with the Monte Carlo expectation 
well within the statistical error band presented in Fig.~\ref{fig1},  no claim of any excess can be made with present statistics.

We also searched for possible detector effects that could be the source of the double line, as speculated in  \cite{Hooper:2012,mirabal},
and do not find any. If there is a systematic detector effect, it must show up not only in the 12 chosen regions but also in the background.
There is no such an effect in the statistically much larger background. We conclude that most probably the observed double structure of the 
9 high-energy photons is just a statistical fluctuation. More data from new experiments with high energy resolution 
should solve the open issue of one peak versus double peak in the observed 130~GeV excess from the Galactic centre and from nearby galaxy 
clusters.

We stress that the results and arguments presented in this note do not affect any way the evidence for 130~GeV excess from Galactic centre and from
nearby galaxy clusters.

{\bf Acknowledgement.}
We thank Doug Finkbeiner for useful communication.
This work was supported by the ESF grants 8090, 8499, 8943, MTT8, MTT59, MTT60, MJD52, MJD272, by the recurrent financing projects SF0690030s09, SF0060067s08 and by the European Union through the European Regional Development Fund.

\end{document}